\documentclass[draft]{agujournal2019}
\usepackage{url} 
\usepackage{lineno}
\usepackage[inline]{trackchanges} 
\usepackage{soul}
\draftfalse

\journalname{Geophysical Research Letters}

\begin{document}

\title{Machine learning for online sea ice bias correction within global ice-ocean simulations}

\authors{William Gregory\affil{1}, Mitchell Bushuk\affil{2}, Yongfei Zhang\affil{1}, Alistair Adcroft\affil{1}, Laure Zanna\affil{3}}

\affiliation{1}{Atmospheric and Oceanic Sciences Program, Princeton University, NJ, USA}
\affiliation{2}{Geophysical Fluid Dynamics Laboratory, NOAA, Princeton, NJ, USA}
\affiliation{3}{Courant Institute of Mathematical Sciences, New York University, New York, NY, USA}

\correspondingauthor{Will Gregory}{wg4031@princeton.edu}

\begin{keypoints} 
\item We use a convolutional neural network (CNN) to perform online sea ice bias correction within global ice-ocean simulations.
\item The CNN systematically reduces the free-running model bias in both the Arctic and Antarctic.
\item The online performance can be improved by combining CNN and data assimilation corrections in order to iteratively augment the training data.
\end{keypoints}


\begin{abstract} 
In this study we perform online sea ice bias correction within a GFDL global ice-ocean model. For this, we use a convolutional neural network (CNN) which was developed in a previous study \cite{Gregory2023} for the purpose of predicting sea ice concentration (SIC) data assimilation (DA) increments. An initial implementation of the CNN shows systematic improvements in SIC biases relative to the free-running model, however large summertime errors remain. We show that these residual errors can be significantly improved with a data augmentation approach, in which sequential CNN and DA corrections are applied to a new simulation over the training period. This then provides a new training data set with which to refine the weights of the initial network. We propose that this machine-learned correction scheme could be utilized for generating improved initial conditions, and also for real-time sea ice bias correction within seasonal-to-subseasonal sea ice forecasts.
\end{abstract}

\section*{Plain Language Summary} 
Climate models contain errors which often lead to predictions which are consistently out of agreement with what we observe in reality. In some cases we know the origin of these errors, for example predicting too much sea ice as a result of consistently cool ocean temperatures. In reality however, there are typically numerous model errors interacting across the atmosphere, ocean and sea ice, and to manually parse through large volumes of climate model data in an attempt to isolate these errors in time and space is highly impractical. Machine learning on the other hand is a framework which is well-suited to this task. In this work we take a machine learning model which, at any given moment, ingests information about a climate model's atmosphere, ocean and sea ice conditions, and predicts how much error there is in the climate model's representation of sea ice, without seeing any actual sea ice observations. We use this to adjust the sea ice conditions in one particular climate model as it is running forward in time making predictions, and we find that this significantly reduces the model's sea ice errors globally.

\section{Introduction}
Machine learning (ML) algorithms are beginning to cement their position as viable subgrid-scale climate model parameterizations, through their ability to isolate complex non-linear relationships within large volumes of high dimensional data \cite{Brenowitz2018,Gentine2018,OGorman2018,Yuval2020,Finn2023,Sane2023}. Typically this is achieved by training an ML model to learn a functional mapping which characterizes the impact of subgrid processes on resolved scales, by training on high resolution simulations or observational data. Significant effort is currently being afforded to the development of these ML parameterizations in the context of e.g., ocean turbulence, with early results \cite{Zanna2020,Frezat2022,Ross2023,Kurz2023,Zhang2023} highlighting their potential to improve important climate statistics, such as eddy kinetic energy at large scales, over their traditional physics-based counterparts.

Alternatively, combining data assimilation (DA) and ML has shown to be a promising framework for learning either subgrid parameterizations or systematic model errors across various domains \cite{Bonavita2020,Brajard2021,Farchi2021,Mojgani2022,Chen2022,Laloyaux2022,He2023}. In a recent study by \citeA{Gregory2023}, hereafter G23, the authors presented a DA-based ML framework in which convolutional neural networks (CNNs) were used to predict state-dependent sea ice errors within an ice-ocean configuration of the Geophysical Fluid Dynamics Laboratory (GFDL) Seamless system for Prediction and EArth System Research (SPEAR) model, as a way to highlight the feasibility of a data-driven sea ice model parameterization within SPEAR. They approached this by first showing that the climatological sea ice concentration analysis increments from an ice-ocean DA experiment map closely onto the systematic bias patterns of the equivalent free-running model. This suggested that an ML model which is able to predict the analysis increments could, in principle, reduce sea ice biases as an online model parameterization or bias correction tool. Their subsequent CNN architecture then used information from local model state variables and their tendencies, to make predictions of the corresponding sea ice concentration analysis increment at any grid cell location. These offline predictions were shown to generalize well to both the Arctic and Antarctic domains, and across all seasons. However, offline performance does not always directly translate to online simulations, which can sometimes exhibit instabilities as well as climate drift after implementation \cite{Rasp2018,Ott2020,Brenowitz2020}. In such cases, the ML model may require an additional online training step in order to sample a larger model state space to which it was initially trained \cite{Rasp2020}. 

In this present work, we advance the field of ML-based parameterizations by investigating the online performance of the G23 DA-based ML model when used as a tool to correct short-term sea ice error growth. We implement the correction scheme here within a coupled ice-ocean configuration of SPEAR, as the G23 CNN was originally trained on data from an ice-ocean DA system, which therefore allows us to make direct comparisons of model biases and increments produced from both the CNN and DA simulations. If the CNN is able to reduce sea ice biases relative to the free-running model, then this will provide a solid foundation for future work into assessing the generalization to fully coupled systems, and ultimately a physics-based sea ice model parameterization.

\section{Data and methods}
\subsection{SPEAR ice-ocean model} 
SPEAR is a fully coupled ice-ocean-atmosphere-land model \cite{Delworth2020}, which shares the same components as the GFDL CM4 model \cite{Held2019}, however with parameterizations and resolutions geared toward seasonal-to-decadal prediction. The ocean and sea ice components are configured at a 1$^\circ$ horizontal resolution and correspond to the Modular Ocean Model v6 (MOM6) and the Sea Ice Simulator v2 (SIS2), respectively \cite{Adcroft2019}. In this work, we consider an ice-ocean configuration of SPEAR, in which MOM6 and SIS2 are forced by atmospheric conditions from the Japanese 55-year Reanalysis for driving ocean-sea-ice models (JRA55-do; \citeA{Tsujino2018}). Details of the ice-ocean experiments are provided in section \ref{sect:IO_exp}.

\begin{figure}[t!]
    \centering
    \includegraphics[width=1\linewidth]{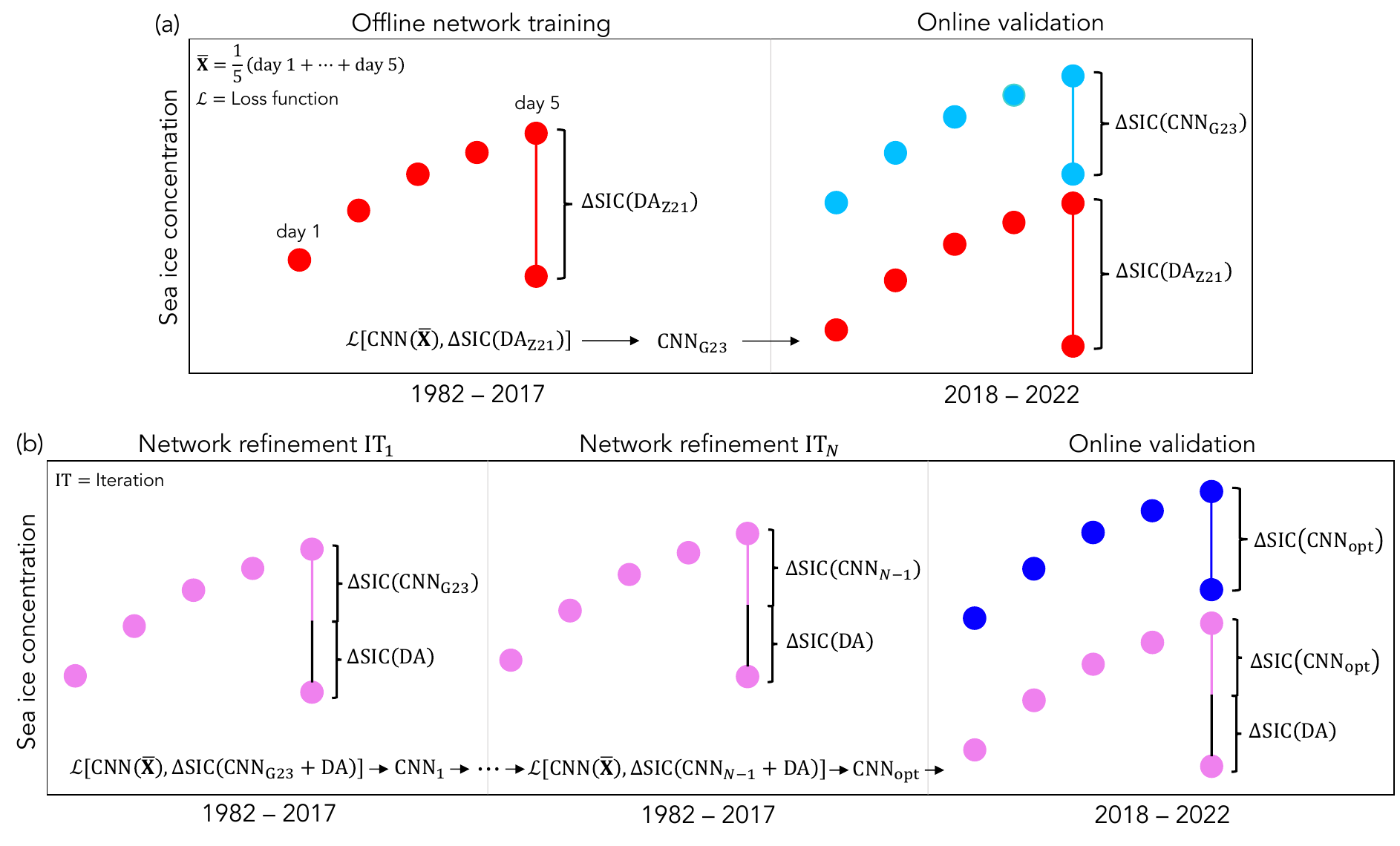}
    \caption{Schematic of bias correction schemes, shown in each panel for one 5-day assimilation/correction cycle. The dots represent the daily model state integrating forward in time. The vertical lines are then the corrections from either the CNN or DA. (a) Out-of-the-box G23 CNN training and implementation showing the Z21 DA simulation in red and the CNN simulation in light blue. (b) Optimized G23 CNN training and implementation showing simulations with combined CNN and DA corrections in violet, and the optimized CNN simulation in dark blue.}
    \label{fig:workflow}
\end{figure}

\subsection{Machine learning model}\label{sect:ML}
The CNN model from G23 was trained to predict sea ice concentration (SIC) increments from a SPEAR ice-ocean DA experiment (\citeA{Zhang2021}; hereafter Z21). The Z21 DA experiment spanned January 1\textsuperscript{st} 1982 -- January  1\textsuperscript{st} 2018, where satellite observations of SIC from the National Snow and Ice Data Center (NSIDC; \citeA{Cavalieri1996}) NASA Team algorithm were assimilated into SIS2 every 5 days using the Ensemble Adjustment Kalman Filter (EAKF) approach \cite{Anderson2001}, and sea-surface temperatures were nudged towards observations from version 2 of the Optimum Interpolation Sea-Surface Temperature (OISSTv2) data set \cite{Reynolds2007,Banzon2016} at the model timestep. It should be noted that SIS2 has a 5-category ice thickness distribution \cite{Bitz2001}, with lower thickness bounds of 0.0, 0.1, 0.3, 0.7, and 1.1 meters. The (observable) aggregate SIC field is therefore computed in the model as the sum of the sea ice concentration in each category (SICN), hence SIC $=\sum_{k=1}^5$SICN$_k$. Similarly, we compute the aggregate SIC increment ($\Delta$SIC) as the sum of the analysis increments in each category ($\Delta$SICN). The G23 CNN then uses 5-day mean inputs of state variables and tendencies corresponding to: SIC, sea-surface temperature (SST), zonal and meridional components of ice velocities (SIU and SIV, respectively), sea ice thickness (SIT), net shortwave radiation (SW), ice-surface skin temperature (TS), sea-surface salinity (SSS), and a land-sea mask, in order to predict $\Delta$SIC. This prediction of $\Delta$SIC is then passed to a second CNN, along with SICN, to predict the category concentration increments $\Delta$SICN. For convenience we refer to these two CNNs as a single network hereafter.

The implementation of the CNN into SIS2 here is performed in an analogous manner to DA. Specifically, we run an ensemble forecast of the model for 5 days (e.g., from 00:00 hours UTC on January 1\textsuperscript{st} to 00:00 UTC on January 6\textsuperscript{th}), where we then generate the corresponding $\Delta$SICN predictions for each ensemble member, add the predicted $\Delta$SICN fields to the instantaneous SICN state (i.e., the state at 00:00 UTC on January 6\textsuperscript{th}), and restart the model for the next 5-day forecast (schematics of this 5-day forecast plus correction process are shown in Figure \ref{fig:workflow}, although section \ref{sect:IO_exp} describes this figure in more detail). It is important to note that we also apply a post-processing after each correction. For this we follow the Z21 procedure for updating sea ice variables during DA, which is as follows: first we remove non-physical values from the updated SICN terms by applying a lower bound of 0 to each category, and then scaling each category by 1/SIC if the updated SIC is greater than 1. Secondly, in the case where the correction is removing all sea ice within a given grid cell, we set the corresponding ice and snow thickness, enthalpy, and ice salinity to 0, and subsequently set the ice-surface skin temperature to the freezing point of sea water, $-1.8^\circ$C. In the case where the correction is adding sea ice to a given category which was previously ice-free, we set the thickness of the ice to the mid-point value within the ice thickness distribution bounds (given as 0.05, 0.2, 0.5, 0.9, 1.3 meters). We then set the salinity, enthalpy and skin temperature of the ice to 5 psu, $-87576$ J, and $-0.36^\circ$C, respectively (conditions based on an initial liquid fraction of frazil ice of 0.75). We also ensure that the newly added sea ice contains no overlying snow.

\subsection{Ice-ocean experiments}\label{sect:IO_exp}
We compare four ice-ocean simulations in this study, where each extends for a 5-year period between January 1\textsuperscript{st} 2018 and January 1\textsuperscript{st} 2023. The initial ice and ocean conditions for all simulations are based on those from the Z21 DA experiment, which ended January 1\textsuperscript{st} 2018. The atmospheric forcing is provided by JRA55-do reanalysis version 1.5, SSTs are nudged towards OISSTv2 observations using a piston velocity of 4 meters per day, and SSS is nudged to a seasonal climatology with a piston velocity of 1/6 meters per day. The experiments are given as follows: 
\begin{enumerate}
\item The free-running model in ice-ocean mode (FREE). 
\item An extension of the ice-ocean DA experiment (DA$_\mathrm{Z21}$), which serves as the benchmark for this study.
\item An `out-of-the-box' implementation of the G23 network (CNN$_\mathrm{G23}$), where the network has been trained offline using all available data from the original DA experiment. This procedure is highlighted in Figure \ref{fig:workflow}a, where, during training, the loss function $\mathcal{L}$ minimizes the error between the network predictions, CNN($\bar{\mathbf{X}}$), and the increment from DA, $\Delta$SIC(DA$_\mathrm{Z21}$). Here $\bar{\mathbf{X}}$ represents the 5-day mean state variables and tendencies described in section \ref{sect:ML}. After training, this produces the network CNN$_\mathrm{G23}$, which is then implemented over the 2018--2022 period. The reader is referred to G23 for more details of the architecture and hyperparameters related to the network training process.
\item An `optimized' version of the G23 network (CNN$_\mathrm{opt}$) where the weights of the G23 network are refined to improve online performance. For this we use both DA and CNNs to iteratively augment the training data, and subsequently refine the network weights after each augmentation iteration. For example, in the first iteration we run a new ice-ocean simulation between 1982--2017, and in which we apply a two-step CNN+DA correction every 5 days; first using CNN$_\mathrm{G23}$, and then using DA (see Figure \ref{fig:workflow}b). We then use this 36-year simulation as a new training data set with which to update the weights of CNN$_\mathrm{G23}$, where, during training, the loss function now minimizes the error between the network predictions, CNN($\bar{\mathbf{X}}$), and the total model error, $\Delta$SIC(CNN$_\mathrm{G23}$+DA). This procedure is performed for a total of $N=3$ iterations. The network refinement after each augmentation iteration is performed in an identical way to the offline learning procedure outlined in G23, except now we only update the weights for 5 epochs after each iteration. 
\end{enumerate}
Note that the `Online validation' panels in Figure \ref{fig:workflow} highlight the simulations with CNN implementations relative to a simulation which applies the respective `perfect' correction (i.e., the correction which either CNN would produce if it had 100\% prediction accuracy). These simply correspond to the extended DA experiment for CNN$_\mathrm{G23}$ in Figure \ref{fig:workflow}a, and the two-step CNN correction plus DA for CNN$_\mathrm{opt}$ in Figure \ref{fig:workflow}b. A comparison of these corrections (increments) is made in section \ref{sect:increments} in order to establish how the different implementation configurations manifest within the increments. As a final point to note here, all results presented in this work are based on ensemble mean fields, and all simulations are run with a `no leap' calendar, which excludes leap-year days.

\begin{figure}[t!]
    \centering
    \includegraphics[width=1\linewidth]{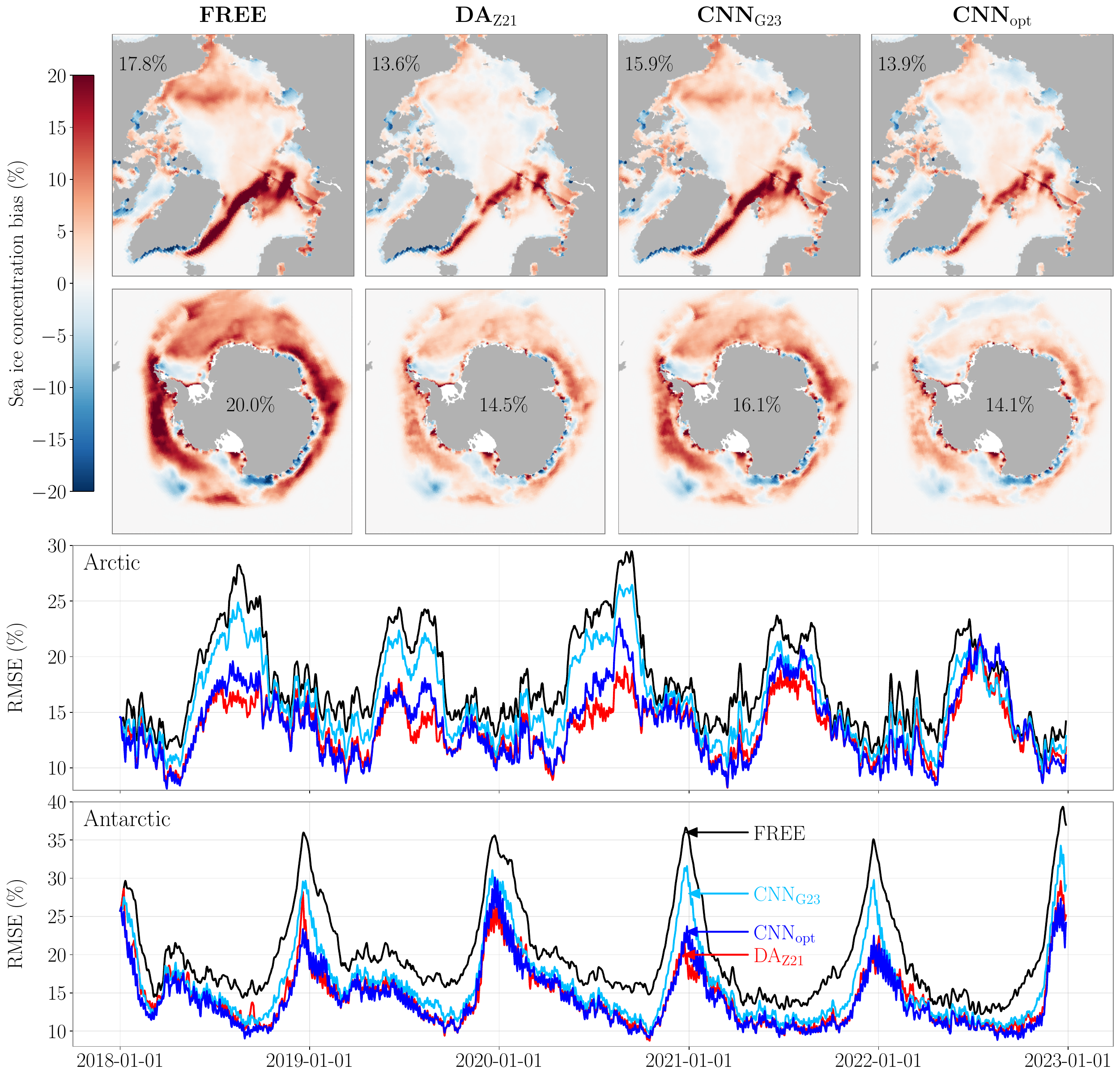}
    \caption{Comparison of model biases from ice-ocean simulations over the 2018--2022 period. The first and second rows show the mean SIC biases (model minus observations), relative to the NSIDC NASA Team observations \cite{DiGirolamo2022}. The average RMSE of SIC is reported in each panel (RMSE computed each day over sea ice covered grid cells only, and then averaged over all days). The bottom two time series plots show pan-Arctic and pan-Antarctic RMSE of SIC for each simulation.}
    \label{fig:mainfig}
\end{figure}

\section{Results}
\subsection{Model bias}\label{sect:bias}
Figure \ref{fig:mainfig} shows model biases for each of the ice-ocean experiments outlined in section \ref{sect:IO_exp}. Initially considering the annual-mean spatial bias patterns of the free-running model, we can see that this simulation is overall positively biased in both hemispheres, with largest Arctic biases occurring in the east Atlantic sector (Greenland, Barents, and Kara seas), and largest Antarctic biases in the Bellingshausen, Amundsen, and Indian Ocean sectors. The daily SIC root-mean-squared error (RMSE) curves then highlight the seasonal variation of the model bias, with largest RMSE values in FREE (black curves) occurring across June--August in the Arctic (22.7\%), and December--February in the Antarctic (28.5\%). The average RMSE over the entire simulation period corresponds to 17.8\% and 20.0\% in the Arctic and Antarctic, respectively, with larger errors in summer and smaller errors in winter. As expected, the DA experiment (DA$_\mathrm{Z21}$; red curves) visibly reduces the bias across all seasons, with average Arctic and Antarctic RMSE reductions relative to FREE of 4.2\% and 5.5\%, respectively. 

Turning to the two CNN correction schemes, the out-of-the-box implementation (CNN$_\mathrm{G23}$; light blue curves) shows systematic improvements relative to FREE, with average RMSE reductions of 1.9\% and 3.9\% in the Arctic and Antarctic, respectively. The modest improvements in the Arctic make it difficult to identify qualitative differences in the climatology spatial bias plots, however some improvements can be seen in the east Atlantic sector. This is also highlighted in the regional Arctic sea ice extent (SIE) time series (Figure S1), where regions are defined according to \citeA{Meier2023}. On the other hand, CNN$_\mathrm{G23}$ shows visible improvements across much of the Antarctic domain, with large bias reductions in the Amundsen Sea and Pacific Ocean. The regional Antarctic SIE time series (Figure S2) also more closely track the DA experiment throughout the majority of the simulation period, particularly in the Antarctic growth season. In the melt season however, the simulation shows a tendency to drift back towards to the free-running model state. Comparing these results to the optimized CNN implementation (CNN$_\mathrm{opt}$; dark blue curves), we see marked skill improvements. The average RMSE reductions compared to FREE are 3.9\% and 5.9\% in the Arctic and Antarctic, respectively, and Figure \ref{fig:mainfig} shows that sizeable improvements have been made in the summer months in both hemispheres. Furthermore, both pan-Antarctic and regional SIE (Figure S2) are also considerably improved in the melt season compared to CNN$_\mathrm{G23}$, and often show reduced biases relative to the DA experiment. It is worth noting however that many of the regional Antarctic SIE time series for CNN$_\mathrm{opt}$ show visible imprints of model shock (i.e., large fluctuations in extent, occurring every 5 days). This can occur in DA when there is significant drift between each assimilation cycle, and the fact that we see this here may suggest that there is rapid error growth occurring over the space of 5 days in the Antarctic. We discuss this further in section \ref{sect:increments}. In any case, the fact that the CNN$_\mathrm{opt}$ experiment, which does not assimilate any observations, has similar errors to DA$_\mathrm{Z21}$ suggests that the DA run is primarily correcting systematic model error and that CNN$_\mathrm{opt}$ is successfully capturing these errors. 

\subsection{Understanding online improvements}\label{sect:onOFF}
Between the two CNN models, it is clear that CNN$_\mathrm{opt}$ is the most desirable scheme for reducing the free-running model bias. Furthermore, it is also clear that, relative to CNN$_\mathrm{G23}$,  the largest gains from CNN$_\mathrm{opt}$ come in the summer months. In this section we take a closer look at the performance of each CNN correction scheme in order to discern how these improvements manifest in both the spatial error patterns of each simulation, and also in the SIC increments produced from each scheme.

\subsubsection{Snapshots}\label{sect:snaps}
\begin{figure}[t!]
    \centering
    \includegraphics[width=1\linewidth]{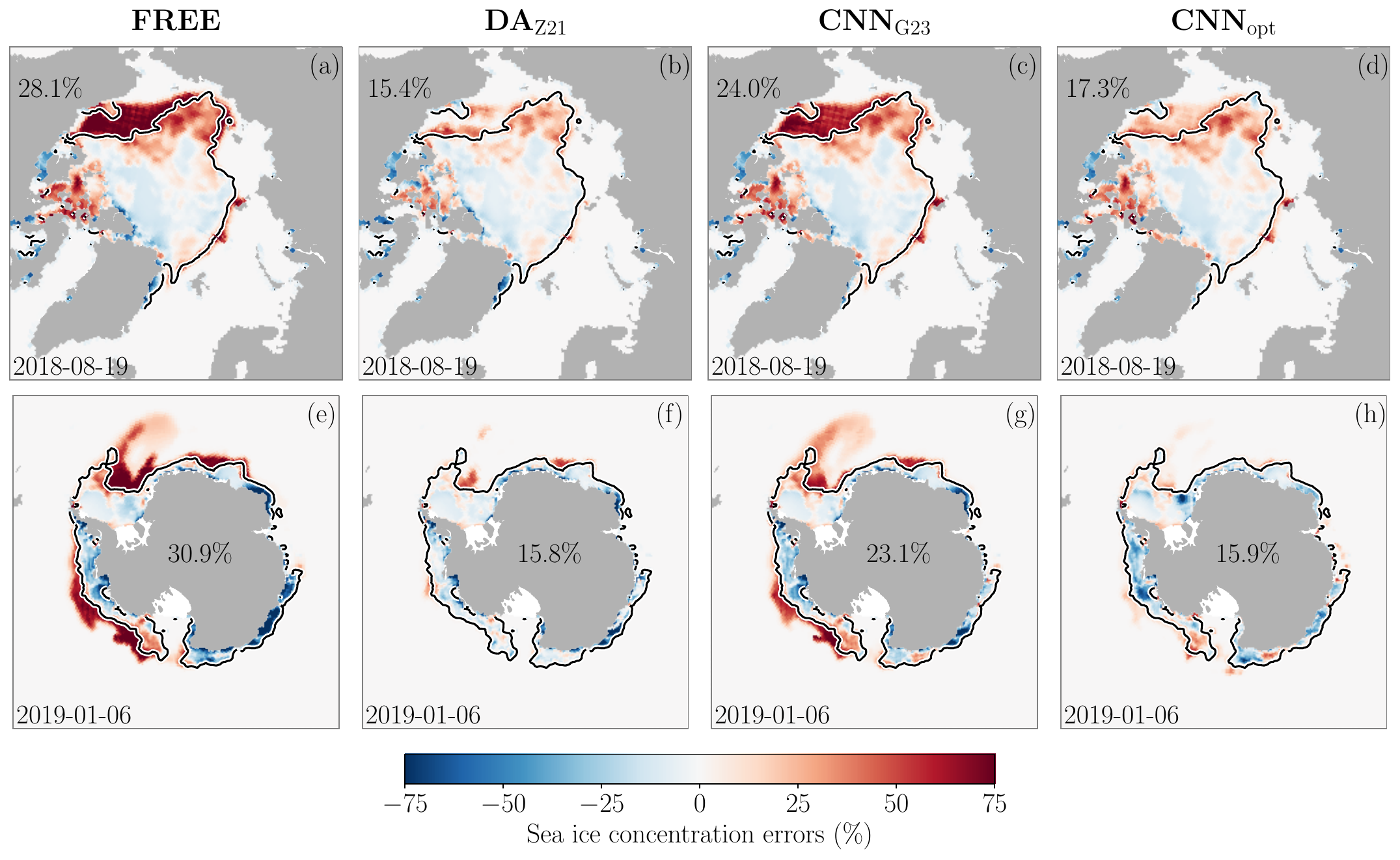}
    \caption{Snapshots of summertime model errors (model minus observations) for the FREE, DA and CNN ice-ocean simulations. Errors are computed relative to the NSIDC NASA Team observations \cite{DiGirolamo2022}. RMSE values are reported in each panel. The black contours mark the observed sea ice edge position (15\% SIC).}
    \label{fig:snapshots}
\end{figure}

Figure \ref{fig:snapshots} shows example snapshots of summertime model errors in each hemisphere (see also supplementary movie S1 for snapshots over the 5-year simulation period). In both hemispheres we can see that FREE (Figures \ref{fig:snapshots}a and \ref{fig:snapshots}e) contains large positive errors related to over-estimation of the sea ice edge (indicated by the positive SIC biases equator-ward of the observed ice edge contour). The errors pole-ward of the ice edge contour indicate local SIC errors. While the DA simulation (Figures \ref{fig:snapshots}b and \ref{fig:snapshots}f) retains some of these local SIC errors, a significant fraction of the ice edge errors are reduced; almost halving the RMSE in the Antarctic relative to FREE. CNN$_\mathrm{G23}$ (Figures \ref{fig:snapshots}c and \ref{fig:snapshots}g) shows some improvements relative to FREE (4.1\% and 7.8\% RMSE improvement in the Arctic and Antarctic, respectively), however there are still considerable ice edge and local SIC errors throughout both the Pacific sector in the Arctic, and the Atlantic and Pacific sectors in the Antarctic. 

For CNN$_\mathrm{opt}$ (Figures \ref{fig:snapshots}d and \ref{fig:snapshots}h) there are clear RMSE improvements relative to both FREE and CNN$_\mathrm{G23}$ in both hemispheres, with remarkable improvements in the Antarctic. It could be argued however that, in the Arctic, the simulated ice edge position is not much improved in this example. A useful metric to confirm this is the integrated ice edge error (IIEE; \citeA{Goessling2016}), which computes the total area for which the ice edge is both over- and under-predicted, relative to satellite observations. For panels (a--d) in Figure \ref{fig:snapshots}, the IIEEs are given as 1.44, 0.90, 1.32, 1.05 million km$^2$, respectively, which shows that the sea ice edge from each correction scheme is in better agreement with the observations than FREE, and that CNN$_\mathrm{opt}$ does indeed improve over CNN$_\mathrm{G23}$ in this regard. Similarly in the Antarctic panels (e--h), the IIEEs are given as 3.98, 1.38, 3.11, 1.30 million km$^2$, respectively. Here we can see that CNN$_\mathrm{opt}$ even shows improved ice edge errors over the DA simulation (see Figure S4 for IIEE metrics computed over the entire 2018--2022 period). We can also take the assessment of ice edge errors further by disaggregating the SIC RMSE metric into grid points that lie pole-ward and equator-ward of the observed ice edge contour on any given day, in order to assess where the largest improvements from each correction scheme are manifesting (i.e., whether improvements are primarily in the ice edge location or SIC within the ice pack). From this decomposition (Figures S5 and S6) we find that, relative to FREE, in both hemispheres the largest RMSE reductions from each correction scheme come from improvements in the ice edge. Furthermore, we find that CNN$_\mathrm{opt}$ is considerably reducing the summer ice edge errors relative to CNN$_\mathrm{G23}$.

\begin{figure}[t!]
    \centering
    \includegraphics[width=1\linewidth]{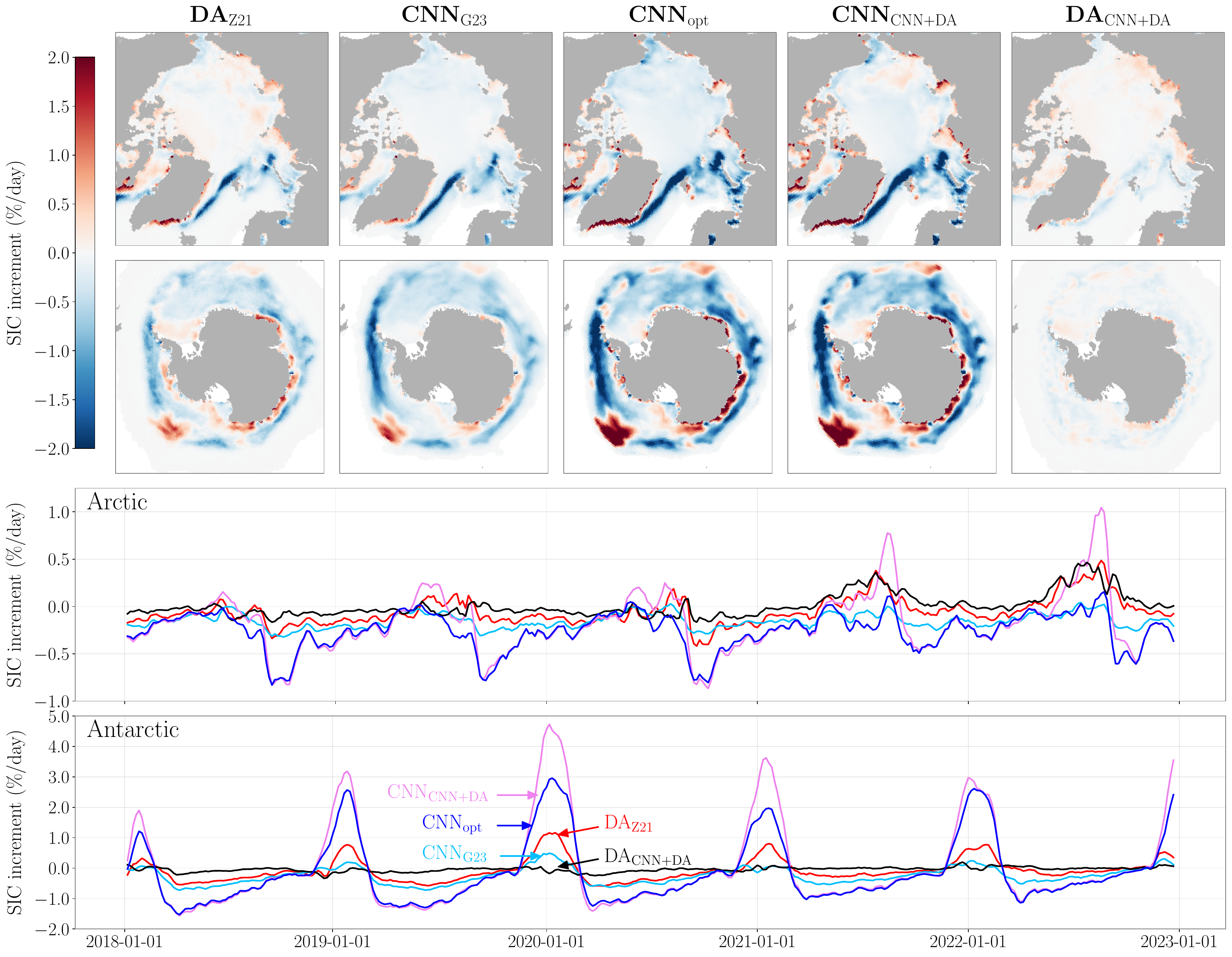}
    \caption{Comparison of SIC increments produced from either DA or CNNs during online simulations. The first two rows show spatial climatologies over 2018--2022. The bottom two panels are then the equivalent time series, computed as mean fields over Arctic and Antarctic domains.}
    \label{fig:increments}
\end{figure}

\subsubsection{Analysis increments}\label{sect:increments}
Figure \ref{fig:increments} shows the mean SIC increments for all DA and CNN correction schemes. The increments here correspond to those which were originally outlined in the validation panels of Figure \ref{fig:workflow}. Namely, the extended DA experiment (DA$_\mathrm{Z21}$), the out-of-the-box CNN implementation (CNN$_\mathrm{G23}$), the optimized version of the G23 network (CNN$_\mathrm{opt}$), and each of the corrections from the two-step CNN$+$DA process, using the optimized network (CNN$_\mathrm{CNN+DA}$ and DA$_\mathrm{CNN+DA}$, respectively).

The mean increments from both CNN$_\mathrm{G23}$ and CNN$_\mathrm{opt}$ in Figure \ref{fig:increments} show largely similar spatial patterns in both hemispheres, with CNN$_\mathrm{opt}$ displaying overall larger magnitudes. In the Arctic, while both sets of CNN increments show isolated regions of positive values along the Eurasian coast, they do not reflect the larger area of mean positive increments seen in DA$_\mathrm{Z21}$ across the East Siberian, Laptev and Kara seas (from the DA$_\mathrm{Z21}$ Arctic time series panel in Figure \ref{fig:increments} we can see that these positive increments originate in summer). On the other hand, the increments from CNN$_\mathrm{CNN+DA}$ do indeed show mean positive summer values in these regions. This suggests that the combination of input variables to the network which are needed to generate these positive predictions in the Arctic, is only being sampled after the additional DA$_\mathrm{CNN+DA}$ step. Nonetheless, in section \ref{sect:snaps} we have seen that CNN$_\mathrm{opt}$ yields significant improvements in Arctic summer SIC errors over CNN$_\mathrm{G23}$, which is primarily coming from improvements in ice edge errors. This is consistent with the larger magnitude negative increments from CNN$_\mathrm{opt}$ in regions such as the Beaufort and Chukchi seas. This may then suggest that the positive summer increments seen in the DA and CNN$_\mathrm{CNN+DA}$ corrections are needed to target local summer SIC errors. Regarding DA$_\mathrm{CNN+DA}$, we can see that, in both hemispheres, these corrections are lower in magnitude than DA$_\mathrm{Z21}$ on average, which highlights how the initial correction from CNN$_\mathrm{CNN+DA}$ is removing a sizeable component of the model error; leaving less error to correct with DA. This is particularly the case in the Antarctic, where the daily increments from DA$_\mathrm{CNN+DA}$ are very close to zero, suggesting that the CNN has effectively removed the systematic component of the model error in the Antarctic. Meanwhile in the Arctic, there are still residual systematic summertime errors associated with under-predicting the positive increments, which DA$_\mathrm{CNN+DA}$ needs to address.

A natural question then arises as to why DA$_\mathrm{Z21}$ is as effective, if not more effective, at reducing the model bias than CNN$_\mathrm{opt}$, even though the increments from CNN$_\mathrm{opt}$ are considerably larger in magnitude. For this we turn to a comparison of the increments from each of the concentration categories (Figures S6 and S7). For CNN$_\mathrm{opt}$, we find that the largest magnitude corrections are being made to the thinnest ice category, while on the other hand, DA$_\mathrm{Z21}$ makes sizeable corrections to some of the thicker categories. This therefore means that CNN$_\mathrm{opt}$ needs to make larger corrections to achieve the same volume change as DA$_\mathrm{Z21}$. Furthermore, the fact that CNN$_\mathrm{opt}$ is largely updating the thinnest ice category also explains the model shock seen in the regional Antarctic SIE time series for CNN$_\mathrm{opt}$ in Figure S2. In the Pacific sector in summer for example, the CNN is adding large extents of ice, which the model is then consistently removing over each 5 day interval. This now seems conceivable given that the new ice is very thin (5 cm thickness), and hence would be susceptible to completely melting if advected to grid cells with sufficiently warm SSTs. Further evidence to support this claim comes from the fact that this model shock behavior is significantly damped in the regional sea ice volume time series (Figure S8); which makes sense as the thinnest ice category will typically contribute less to the regional volume.

\section{Discussion and conclusions}\label{sect:disc}
There is currently much discourse centered around ML-based parameterizations and/or corrections within climate models, particularly in the context of how to achieve stable and unbiased simulations after implementation. Many studies have illustrated how to achieve stability within idealized models, where parameterizations are learned from high resolution simulations. For example, by swapping out neural networks for random forests \cite{Yuval2020,Watt2021}, or using online and/or reinforcement learning \cite{Rasp2020,Kurz2023}.

In this study we have shown that a CNN model which has been trained purely offline to predict increments from a sea ice DA system (which assimilates real sea ice observations) can be used `out-of-the-box' to systematically reduce sea ice biases in a 5-year global ice-ocean simulation, without instabilities or drift. We have also introduced a data augmentation approach to optimize the offline-trained CNN, which significantly improves online generalization in both hemispheres; particularly in terms of reducing sea ice edge errors in summer. This augmentation approach is performed by iteratively generating new simulations in which corrections are applied from both the current iteration of the ML model, as well as DA. Each iteration of the augmentation procedure therefore provides a new training data set with which to refine the CNN weights from the previous iteration. While, in theory, this procedure could be repeated to convergence, we opted for $N=3$ iterations in this study due to computational expense. It is likely however that continued iterations would yield further improvements, particularly in the Arctic summer, given that sizeable gains were made between each of the three iterations here (see Figure S9). We hypothesise that the improvements from this augmentation procedure are a result of exposing the network to input variables which contain information about how the model trajectory evolves after implementing the CNN (as opposed to training purely offline where the inputs have no feedback with the CNN).

Interestingly, we find that, relative to the original DA experiment, the climatological sea ice biases associated with the simulation which uses this `optimized' network are actually modestly improved in the Antarctic (Figure 2). This is understandable when we consider that the target variable during each iteration of the network refinement step is no longer the increment from original DA experiment, but rather the sum of the increments from the two-step CNN$+$DA experiment (recall Figure 1b). Therefore, the model bias from the original DA experiment should not be seen as the lower limit on what is achievable with the CNN. We can see this in Figure S10, where the bias of the simulation which applies this two-step CNN$+$DA procedure is indeed systematically lower than both the original DA experiment and the optimized CNN. This leaves exciting avenues for future work relating to improved initial conditions for numerical prediction. For seasonal predictions with the GFDL SPEAR model for example, initial conditions for the ice and ocean \cite{Zhang2022,Lu2020} are based on DA via Ensemble Kalman filters. The Ensemble Kalman filter is not formally designed to correct for systematic model error, and so this two-step CNN$+$DA procedure could be a way to generate more accurate initial conditions (see Figure S10) with the CNN and DA fixing the systematic and random components of the errors, respectively. Indeed while this is similar to a weak constraint 4-D variational DA approach \cite{Wergen1992,Zupanski1993,Tremolet2007}, the CNN has computational advantages (once trained) in that it does not require the construction of an adjoint model \cite{Bonavita2020,Laloyaux2022}.

Further avenues for future work also include the use of the optimized CNN for making bias corrections to real-time seasonal sea ice forecasts, or sea ice projections on climate timescales. For seasonal prediction, the methodology would follow that which has been presented in this study, except the CNN would be applied within the fully coupled SPEAR model. This would however require several considerations. For example, addressing the issue of model shock seen in the Antarctic (which could potentially be reduced by increasing the frequency of the CNN corrections), and also assessing generalization of the CNN to the fully coupled model, which includes new interactive feedbacks with an atmospheric model. Looking also to longer term climate projections, G23 discussed this in the context of implementing the CNN as a sea ice model parameterization. This would then require further considerations of how to appropriately conserve mass, heat and salt when adding/removing sea ice from the ocean.

\section{Open Research}
\noindent All data for training each CNN are openly available \cite{Gregory2023b}, along with auxiliary data such as the optimized CNN weights and standardization statistics. Python code to pre-process the input data and train the CNNs is also available at the same location.

\acknowledgments
William Gregory, Mitchell Bushuk, Alistair Adcroft and Laure Zanna received M$^2$LInES research funding by the generosity of Eric and Wendy Schmidt by recommendation of the Schmidt Futures program. This work was also intellectually supported by various other members of the M$^2$LInES project, as well as being supported through the provisions of computational resources from the National Oceanic and Atmospheric Administration (NOAA) Geophysical Fluid Dynamics Laboratory (GFDL). We also thank Theresa Morrison and Feiyu Lu for their invaluable feedback on this article.

\bibliography{agusample}

\newpage
\renewcommand{\thefigure}{S\arabic{figure}}
\setcounter{figure}{0}
\title{Supporting Information: Machine learning for online sea ice bias correction within global ice-ocean simulations}

\authors{William Gregory\affil{1}, Mitchell Bushuk\affil{2}, Yongfei Zhang\affil{1}, Alistair Adcroft\affil{1}, Laure Zanna\affil{3}}

\affiliation{1}{Atmospheric and Oceanic Sciences Program, Princeton University, NJ, USA}
\affiliation{2}{Geophysical Fluid Dynamics Laboratory, NOAA, Princeton, NJ, USA}
\affiliation{3}{Courant Institute of Mathematical Sciences, New York University, New York, NY, USA}

\begin{figure}[t!]
    \centering
    \includegraphics[width=1\linewidth]{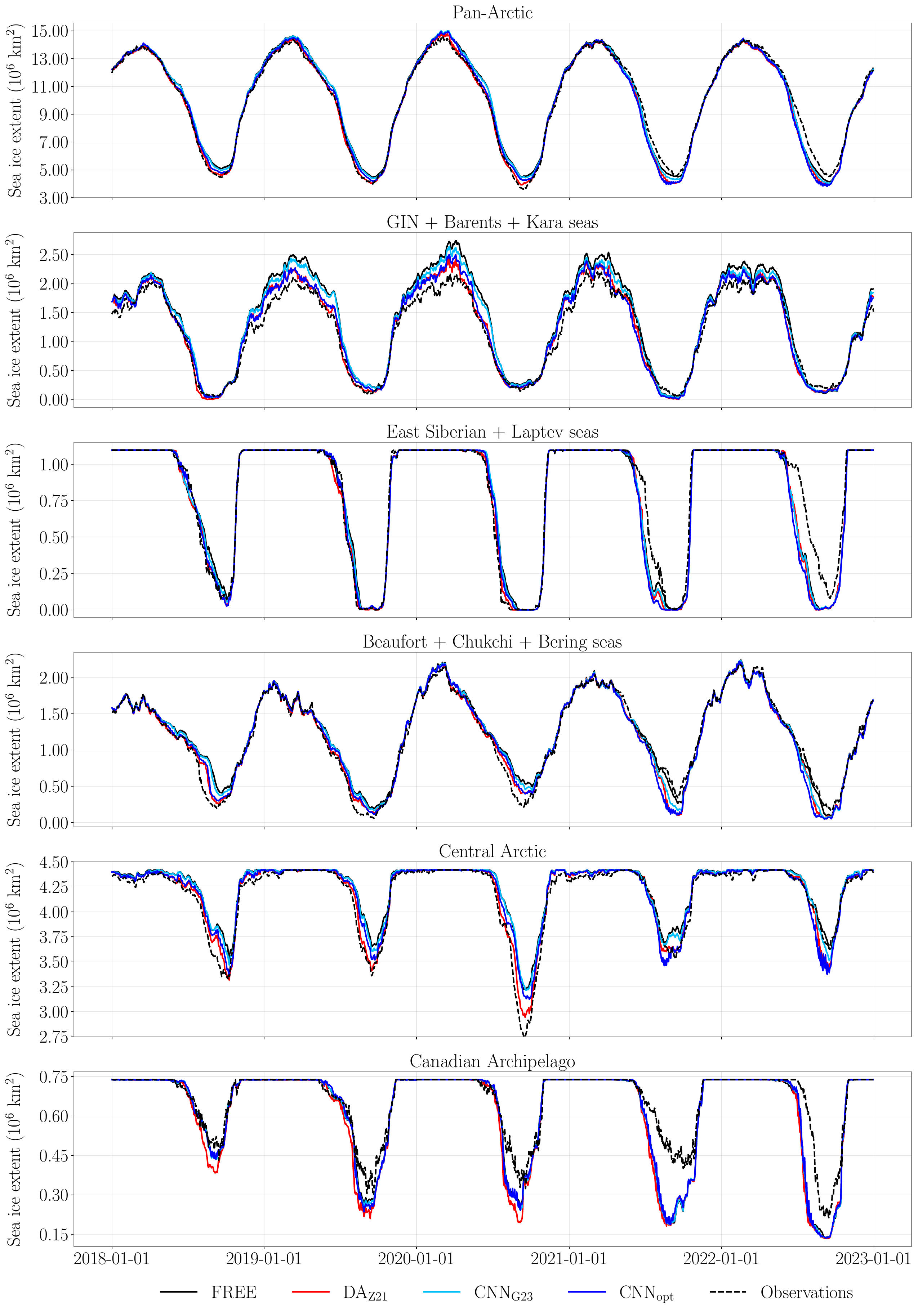}
    \caption{Pan-Arctic and regional daily SIE for ice-ocean experiments over the 2018--2022 period. The experiments correspond to those described in section \ref{sect:IO_exp} of the main article. The observations correspond to SIE computed from the NSIDC NASA Team SIC data set \cite{DiGirolamo2022} The sea ice mask used to compute regional extents was also derived from the NSIDC product \cite{Meier2023}.}
\end{figure}

\begin{figure}[t!]
    \centering
    \includegraphics[width=1\linewidth]{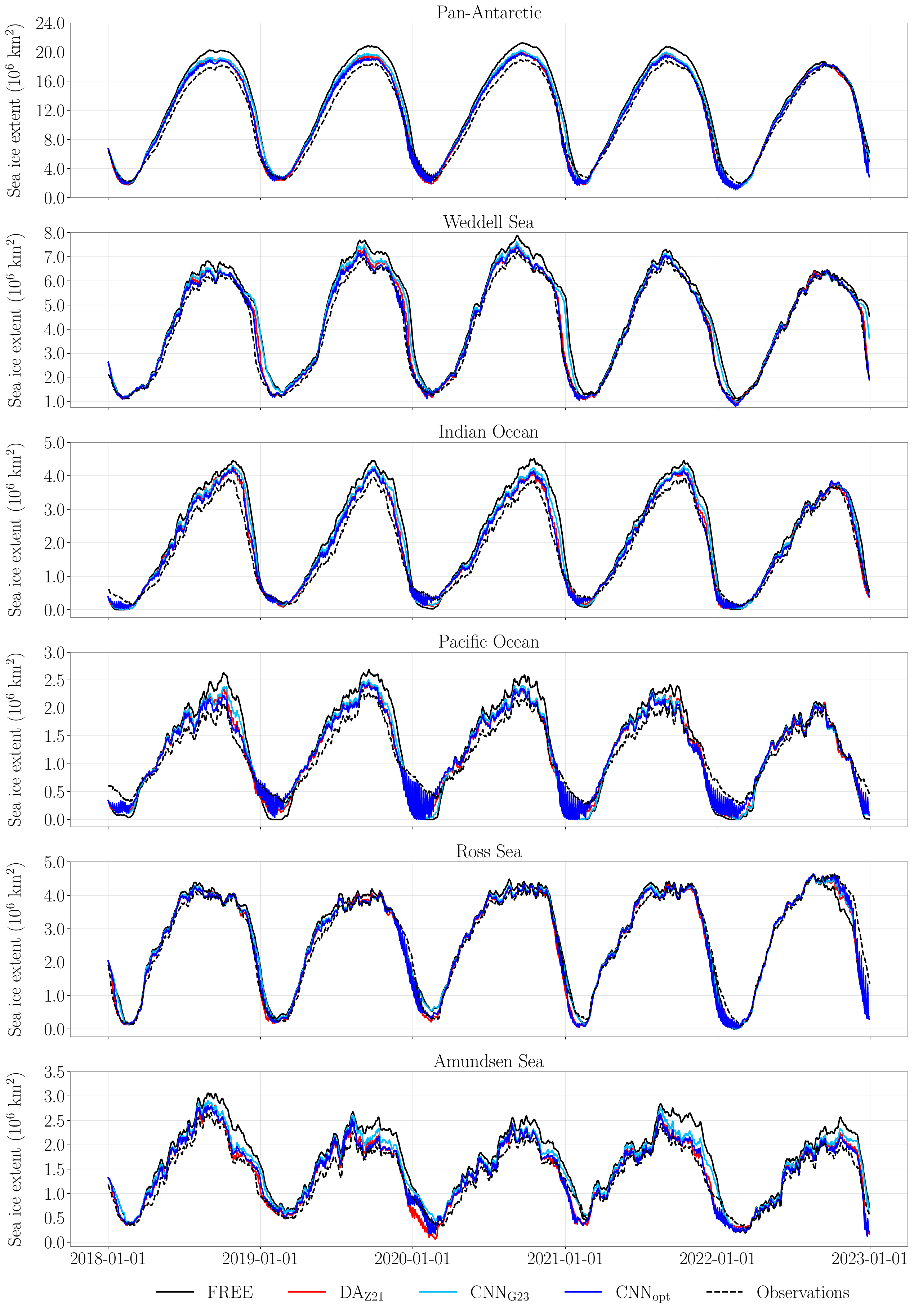}
    \caption{Pan-Antarctic and regional daily SIE for ice-ocean experiments over the 2018--2022 period. The experiments correspond to those described in section \ref{sect:IO_exp} of the main article. The observations correspond to SIE computed from the NSIDC NASA Team SIC data set \cite{DiGirolamo2022} The sea ice mask used to compute regional extents was also derived from the NSIDC product \cite{Meier2023}.}
\end{figure}

\begin{figure}[t!]
    \centering
    \includegraphics[width=1\linewidth]{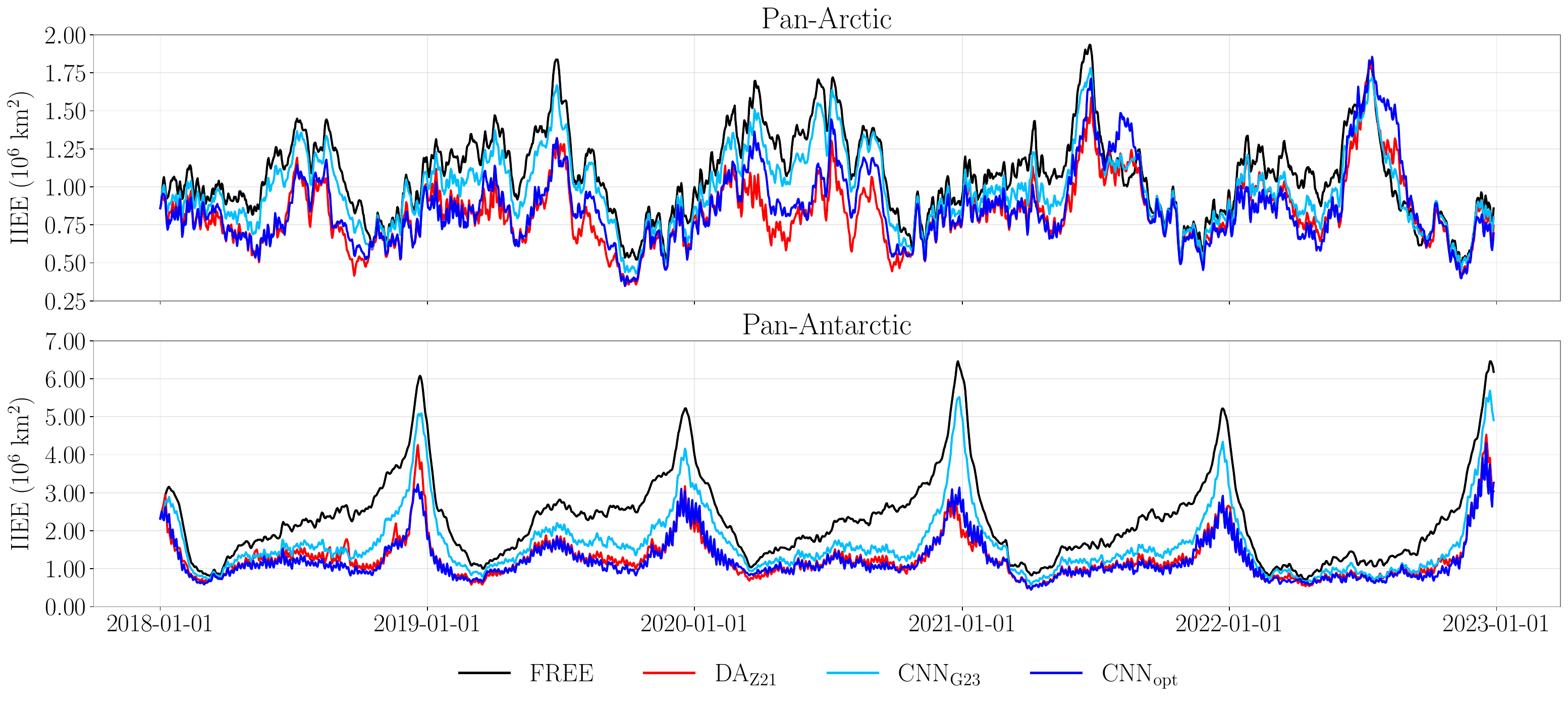}
    \caption{Integrated ice edge error (IIEE) for the various ice-ocean simulations over the 2018--2022 period. IIEE \cite{Goessling2016} quantifies error in the sea ice edge position as the sum of grid cell areas which are either over-predicted or under-predicted. IIEE is computed relative to NSIDC NASA Team SIC observations \cite{DiGirolamo2022}.}
\end{figure}

\begin{figure}[t!]
    \centering
    \includegraphics[width=1\linewidth]{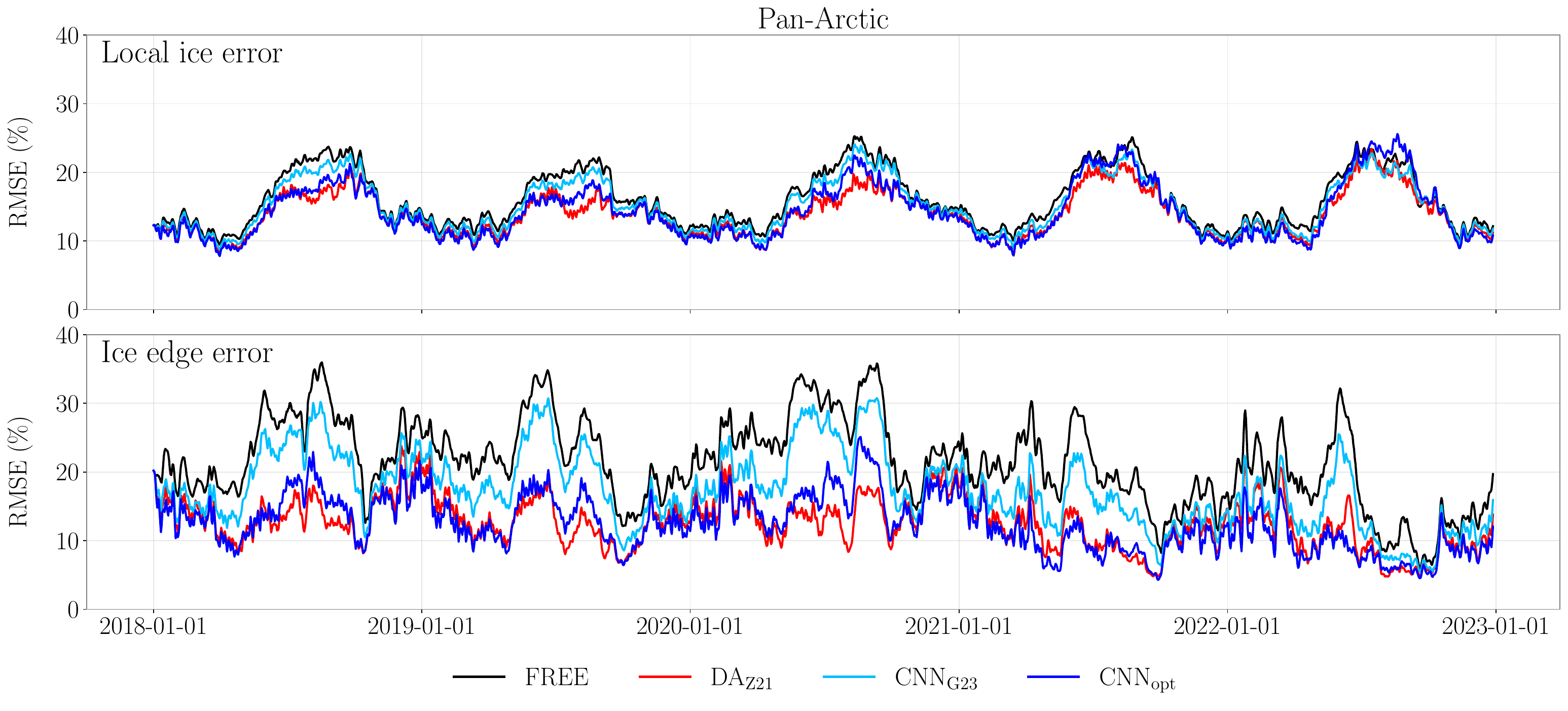}
    \caption{Disaggregation of Arctic RMSE of SIC for CNN simulations, into `local ice errors' (defined as grid points where the observed SIC is greater than or equal to 15\%), and `ice edge errors' (defined as grid points where the observed SIC is less than 15\%). Observations correspond to the NSIDC NASA Team data set \cite{DiGirolamo2022}.}
\end{figure}

\begin{figure}[t!]
    \centering
    \includegraphics[width=1\linewidth]{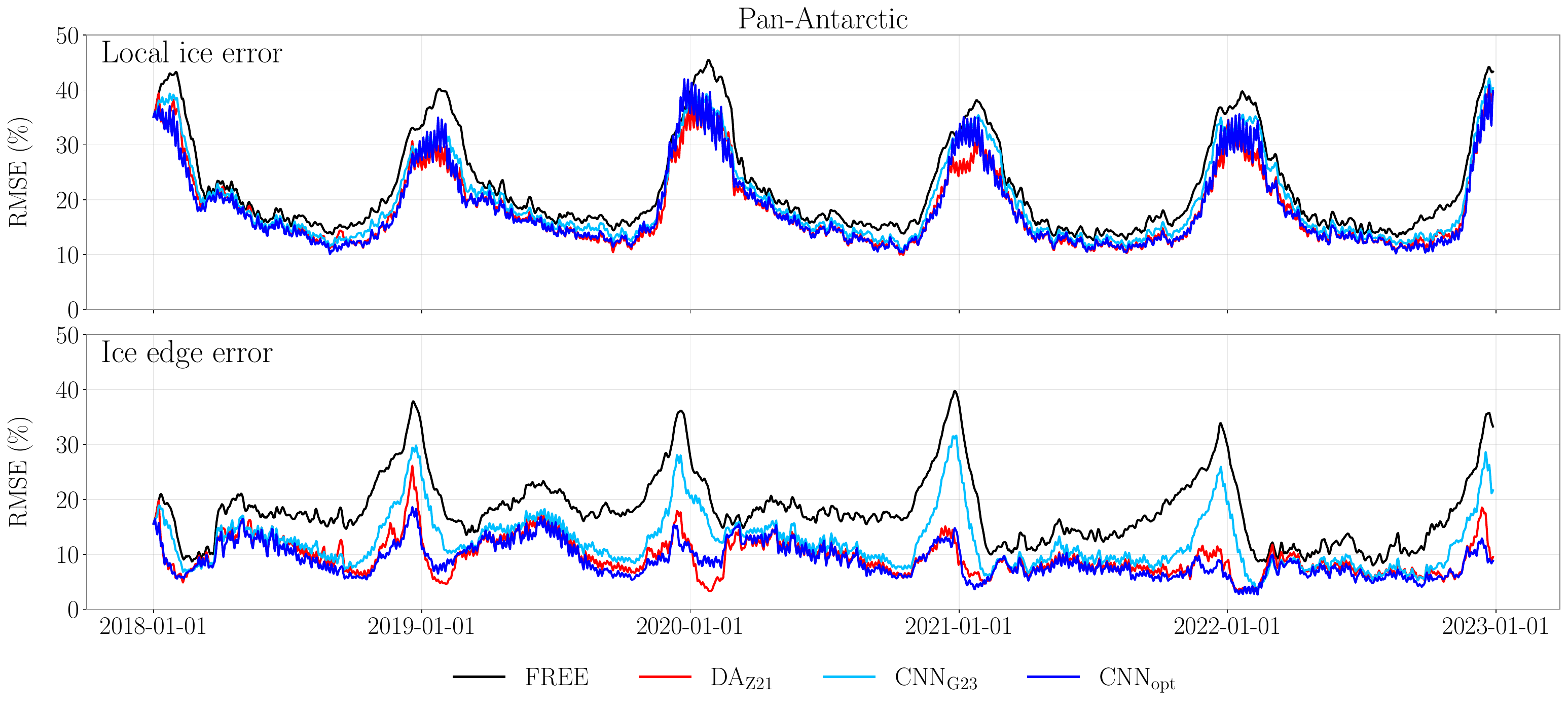}
    \caption{Disaggregation of Antarctic RMSE of SIC for CNN simulations, into `local ice errors' (defined as grid points where the observed SIC is greater than or equal to 15\%), and `ice edge errors' (defined as grid points where the observed SIC is less than 15\%). Observations correspond to the NSIDC NASA Team data set \cite{DiGirolamo2022}.}
\end{figure}

\begin{figure}[t!]
    \centering
    \includegraphics[width=1\linewidth]{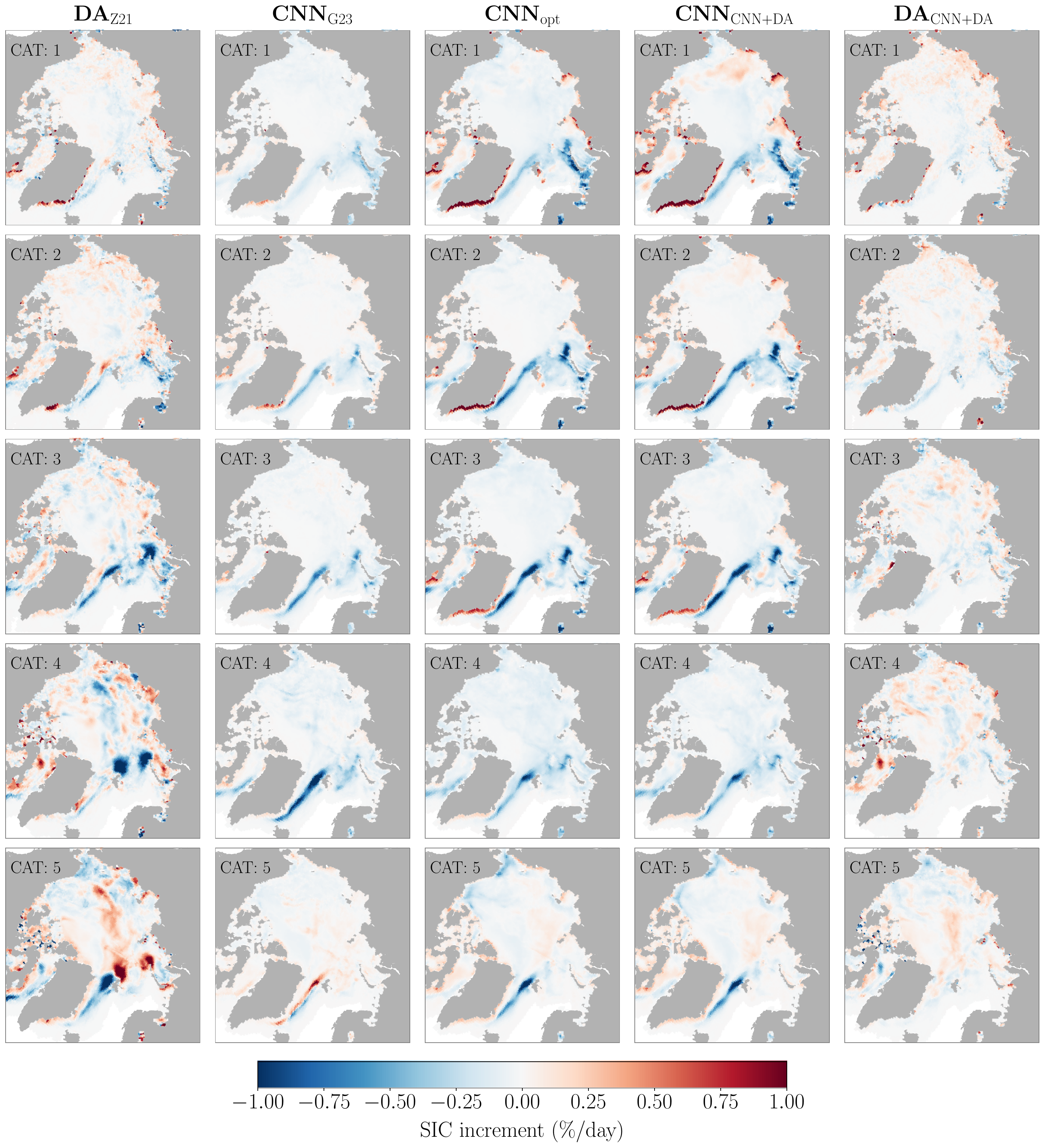}
    \caption{Comparison of category SIC increments in the Arctic, produced from either DA or CNNs during online simulations. Shown as climatologies over 2018--2022. Lower thickness bounds of categories 1 through 5 are: 0.0, 0.1, 0.3, 0.7, and 1.1 meters, respectively.}
\end{figure}

\begin{figure}[t!]
    \centering
    \includegraphics[width=1\linewidth]{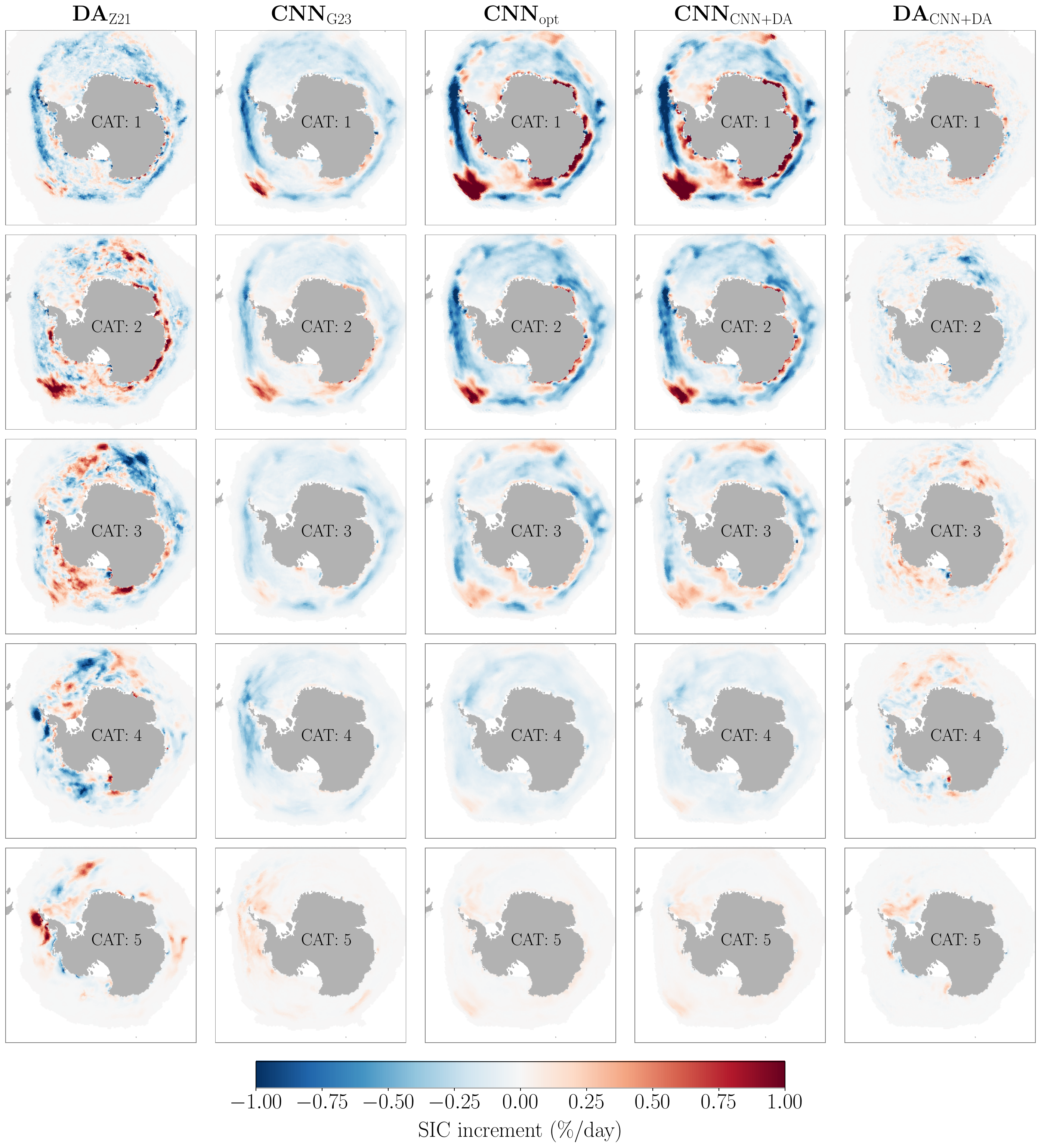}
    \caption{Comparison of category SIC increments in the Antarctic, produced from either DA or CNNs during online simulations. Shown as climatologies over 2018--2022. Lower thickness bounds of categories 1 through 5 are: 0.0, 0.1, 0.3, 0.7, and 1.1 meters, respectively.}
\end{figure}

\begin{figure}[t!]
    \centering
    \includegraphics[width=1\linewidth]{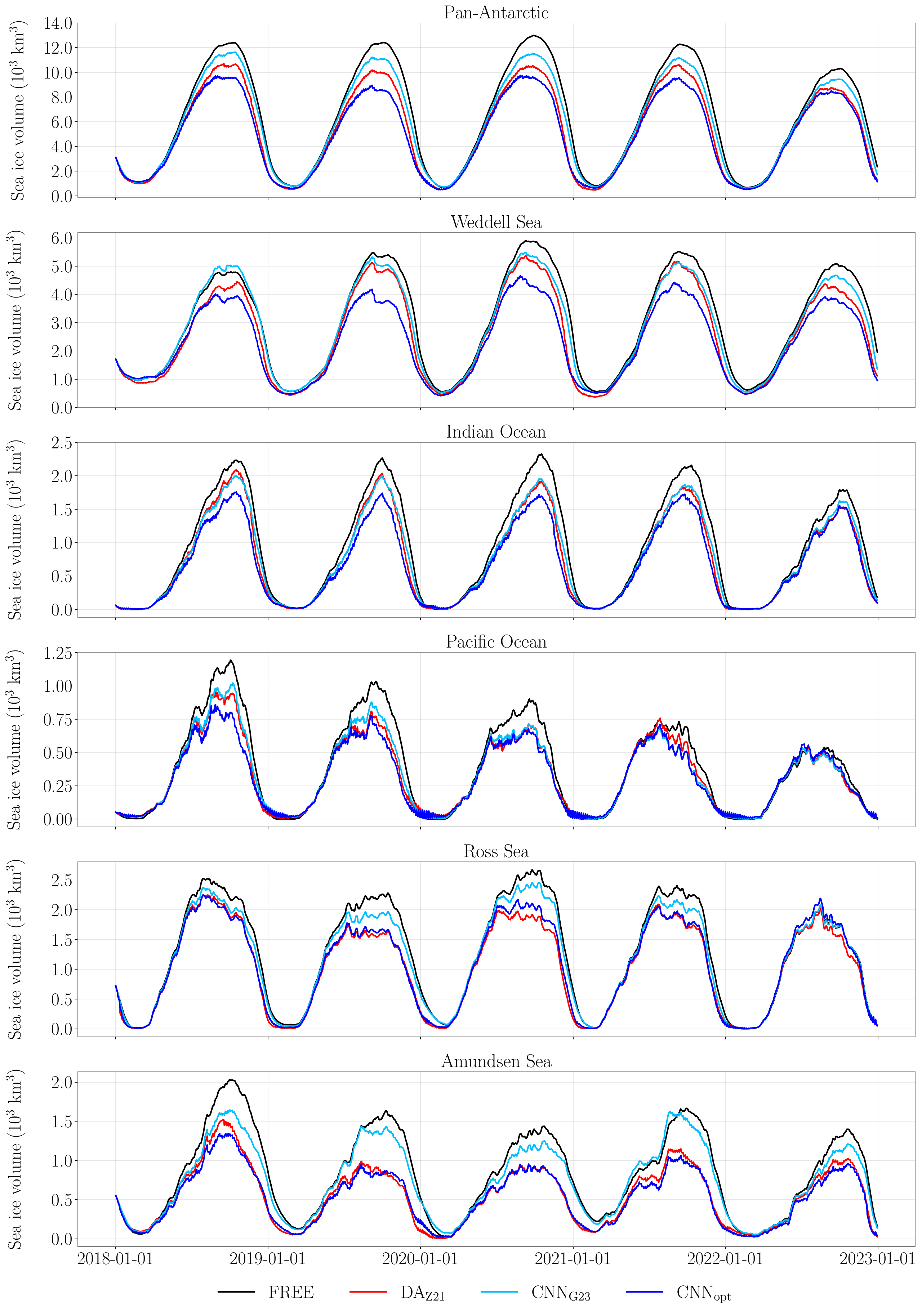}
    \caption{Pan-Antarctic and regional sea ice volume for ice-ocean simulations over the 2018--2022 period. The experiments correspond to those described in section \ref{sect:IO_exp} of the main article. The sea ice mask used to compute regional volume was derived from the NSIDC product \cite{Meier2023}.}
\end{figure}

\begin{figure}[t!]
    \centering
    \includegraphics[width=1\linewidth]{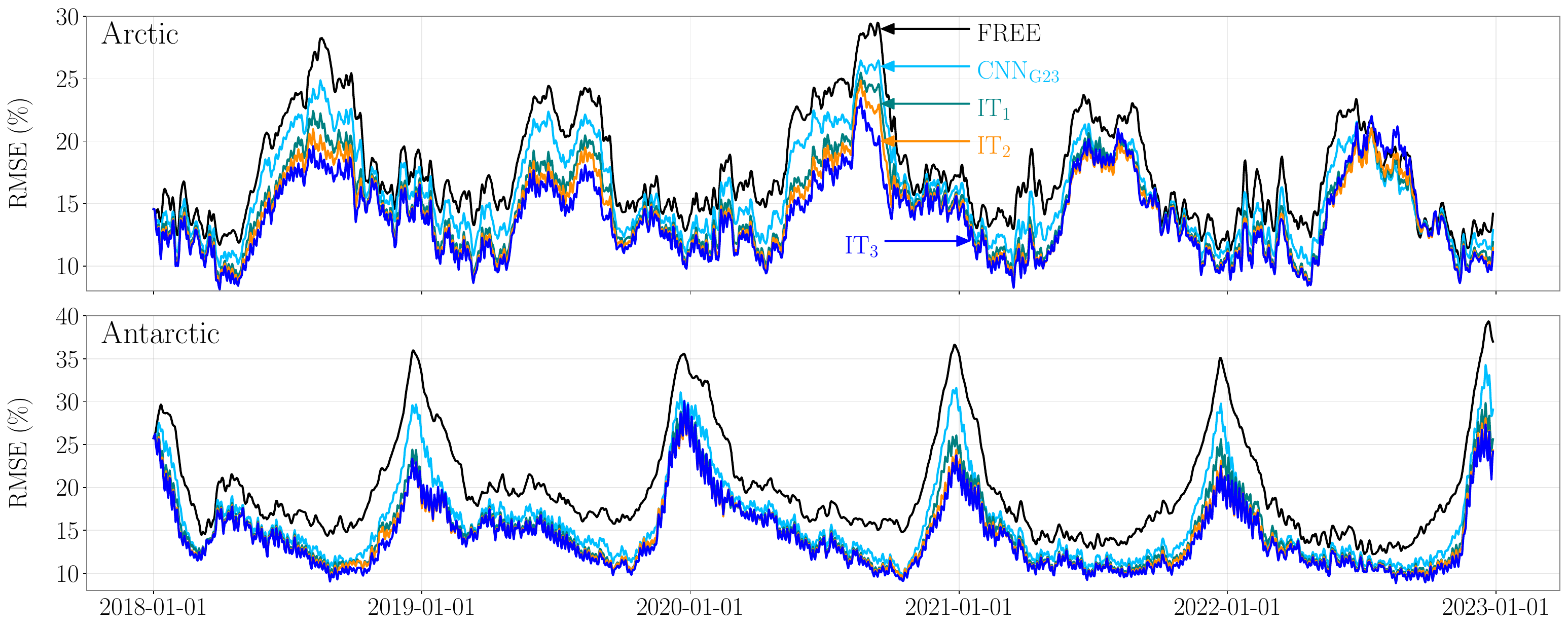}
    \caption{RMSE of SIC for FREE and CNN simulations over 2018--2022. Here the RMSE is shown for each iteration (IT) of the optimized CNN.}
\end{figure}

\begin{figure}[t!]
    \centering
    \includegraphics[width=1\linewidth]{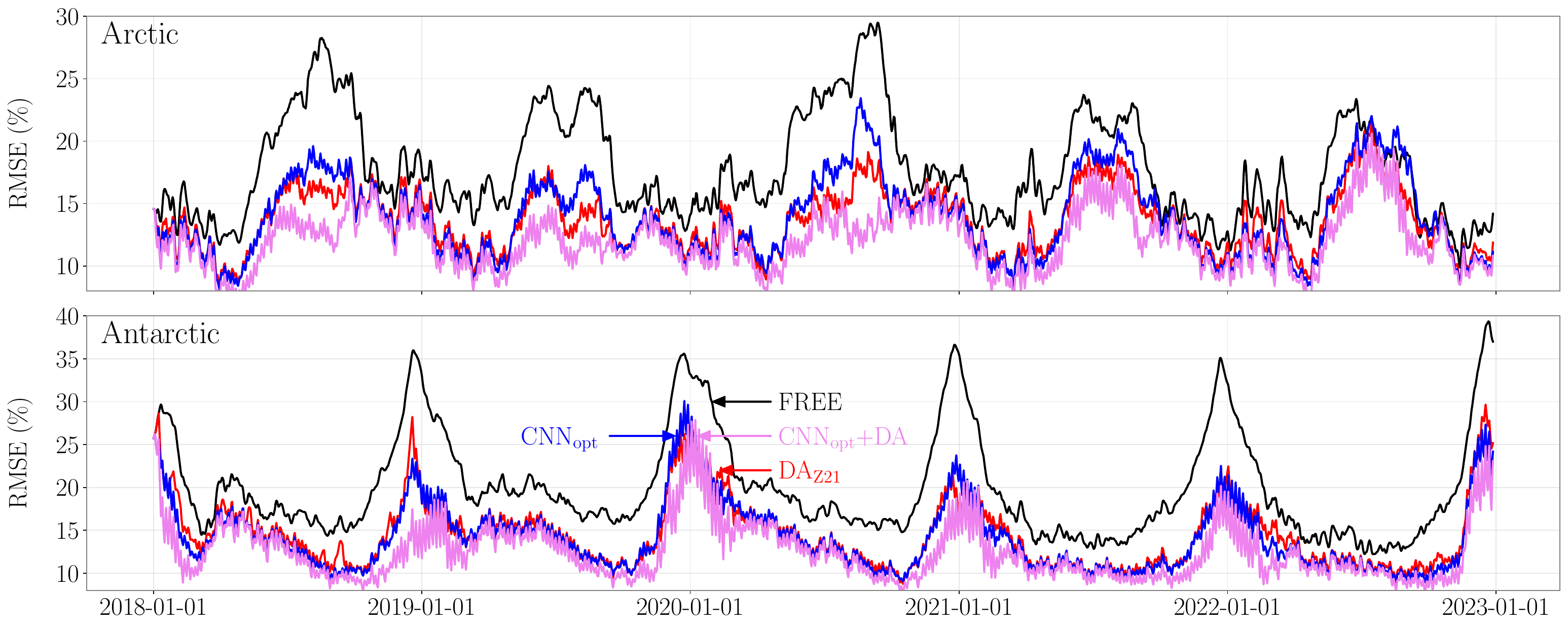}
    \caption{RMSE of SIC for FREE, DA, and CNN+DA simulations over 2018--2022.}
\end{figure}

\end{document}